\documentclass[conference]{IEEEtran}
\IEEEoverridecommandlockouts
\usepackage{booktabs}
\usepackage{multirow}
\usepackage{titlesec}
\usepackage[table,xcdraw]{xcolor} 
\usepackage[linesnumbered,ruled,vlined]{algorithm2e}
\usepackage{amsmath}
\usepackage{tikz}
\usepackage{cite}
\usepackage{amsmath,amssymb,amsfonts}
\usepackage{algorithmic}
\usepackage{graphicx}
\usepackage{textcomp}
\usepackage{xcolor}
\usepackage{url}
\def\BibTeX{{\rm B\kern-.05em{\sc i\kern-.025em b}\kern-.08em
    T\kern-.1667em\lower.7ex\hbox{E}\kern-.125emX}}
\begin{document}

\title{SOME: Symmetric One-Hot Matching Elector \\--- A Lightweight Microsecond Decoder for Quantum Error Correction
}

\author{
Xinyi Guo\textsuperscript{1}, 
Geguang Miao\textsuperscript{2}, 
Shinichi Nishizawa\textsuperscript{3}, 
Hiromitsu Awano\textsuperscript{1}, 
Shinji Kimura\textsuperscript{2}, 
Takashi Sato\textsuperscript{1} \\
\textsuperscript{1}Kyoto University, Kyoto, Japan \\
\textsuperscript{2}Waseda University, Fukuoka, Japan \\
\textsuperscript{3}Hiroshima University, Hiroshima, Japan \\
\IEEEauthorblockA{
guoxinyi@fuji.waseda.jp, geguangmiao@fuji.waseda.jp, 
nishizawa@hiroshima-u.ac.jp,\\
awano@i.kyoto-u.ac.jp, shinji\_kimura@waseda.jp, takashi@i.kyoto-u.ac.jp}
}

\maketitle

\begin{abstract}
Conventional quantum error correction (QEC) decoders such as Minimum-Weight Perfect Matching (MWPM) and Union-Find (UF) offer high thresholds and fast decoding, respectively, but both suffer from high topological complexity. In contrast, Ising model-based decoders reduce topological complexity but demand considerable decoding time.
We propose the Symmetric One-Hot Matching Elector (SOME), a novel decoder that reformulates the QEC decoding task as a Quadratic Unconstrained Binary Optimization (QUBO) problem---termed the One-Hot QUBO (OHQ). Each variable in the QUBO represents whether a given pair of flipped syndromes is matched, while the error probabilities between the pair are encoded as interaction coefficients (weight). Constraints ensure that each flipped syndrome is matched exactly once. Valid solutions of OHQ correspond to self-inverse permutation matrices, characterized by symmetric one-hot encoding. 
To solve the OHQ efficiently, SOME reformulates the decoding task as the construction of permutation matrices that minimize the total weight. It initializes each candidate matrix from one of the minimum-weight syndrome pairs, then iteratively appends additional pairs in ascending order of weight, and finally selects the permutation matrix with the lowest total energy.
SOME achieves up to a 99.9x reduction in variable count and reduces decoding times from milliseconds to microseconds on a single-threaded commodity CPU. OHQ also maintains performance up to a 10.5\% physical error rate, surpassing the highest known threshold of MWPM\@.

\end{abstract}


\section{Introduction}
Quantum computers establish a new computational paradigm, taking advantage of quantum mechanics and quantum effects, to address problems considered intractable for classical computers, enabling solutions in polynomial time~\cite{nielsen2002quantum, hoefler2023disentangling, marella2020introduction}. However, a major challenge in the realization of practical quantum computers is that qubits are highly susceptible to noise and decoherence, leading to unavoidable errors that can corrupt stored information~\cite{kumar2022securing, orszag2010coherence}. Therefore, ensuring reliable quantum computation requires fault-tolerant quantum computing (FTQC), wherein quantum error correction (QEC) plays a crucial role in identifying and correcting such errors~\cite{preskill1998fault, gaitan2008quantum}.

In QEC, there are two key components: one is encoding a ``code" and the other is ``decoding" the code. ``Code" plays a crucial role as it enhances resistance to quantum errors through redundancy~\cite{shor1995scheme}, where multiple physical qubits are combined to form a single logical qubit. Among various coding schemes, Kitaev’s surface code~\cite{kitaev2003fault} and its variants~\cite{fowler2012surface, barends2014superconducting} are regarded as the leading candidates due to their robust error tolerance and compatibility with scalable qubit architectures, as they rely exclusively on local gate operations between neighboring qubits. The surface code consists of two types of qubits: data qubits, which store quantum information, and ancilla qubits, which detect errors.

\begin{figure*}
	\centering
	\includegraphics[width=1.\linewidth]{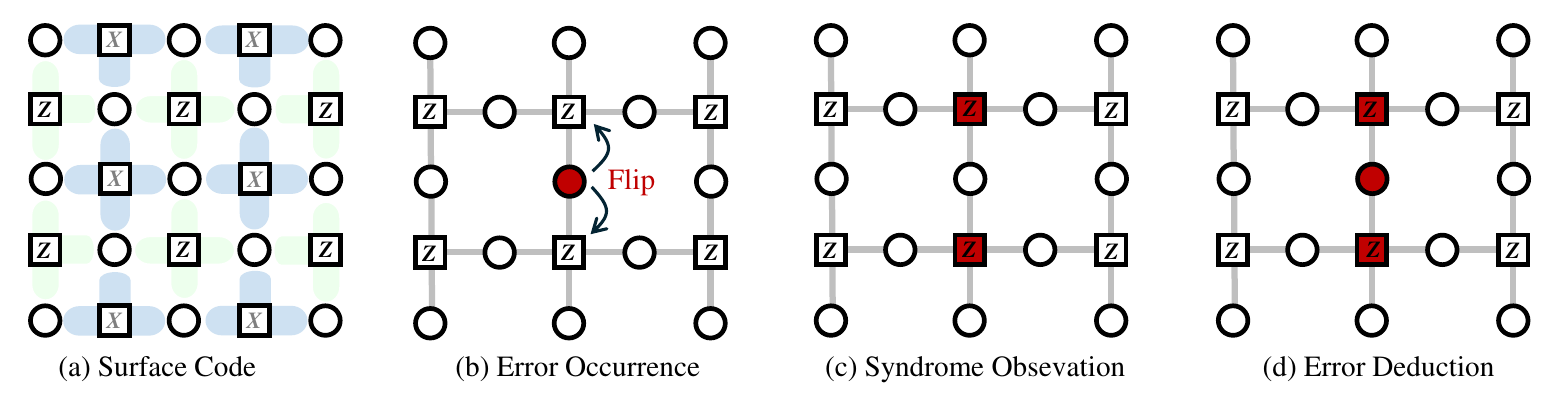}
	\caption{(a) Planar surface code with code distance $d=3$. Open circles indicate data qubits, while squares denote X and Z ancilla qubits. (b)--(d) depict the cases with only X errors. An X error on a data qubit is represented as a red-filled circle~(b), flipping the syndrome from $+1$ to $-1$, thereby turning the open squares into filled red squares~(c). By analyzing these Z syndromes measured by the Z ancilla qubits, the most-likely location of X error can be identified~(d).}
	\label{fig:qec}
\end{figure*}

``Decoding" is essential for identifying which data qubits have experienced errors, based on the measurements from the ancilla qubits, known as syndromes. Effective QEC decoders must operate quickly enough to prevent error accumulation~\cite{terhal2015quantum}, such as on the microsecond timescale for superconducting transmon qubits~\cite{google2023suppressing, marques2022logical}, and on the millisecond timescale for trapped ions~\cite{ryan2021realization}, while also achieving sufficient scalability to correct codes with larger code distances, providing greater redundancy leading to a desirable logical error rate~\cite{google2023suppressing}.

Many QEC decoders have been developed, with minimum-weight perfect matching (MWPM) being a prominent approach~\cite{galil1986efficient, vittal2023astrea, higgott2022pymatching}. This method reformulates the decoding problem as a graph-matching task but is constrained by its worst-case time complexity, which ranges from $O(n^3)$ and $O(n^7)$ depending on the number of data qubits $n$. As a result, real-time decoding is feasible only for code distances up to $d = 9$~\cite{vittal2023astrea}. By leveraging parallel processing, the complexity can be reduced to alomost linear-time~\cite{fowler2012towards, higgott2025sparse, 10313859}, allowing fast MWPM to handle code distances up to $d=33$ under a circuit-level noise of 0.1\%~\cite{10313859}. However, MWPM-based decoders require significant memory resources, leading to considerable overhead for practical implementations~\cite{azad2020distributed}.

Beyond the MWPM decoder, several other classical QEC decoding approaches have been explored. The tensor network decoder~\cite{bravyi2014efficient} reformulates decoding as a machine learning problem, achieving a time complexity of $O(n^2)$. Neural network decoders~\cite{ueno2022neo, varsamopoulos2019comparing, gicev2023scalable} have also demonstrated potential, though their effectiveness depends on extensive training, requiring significant computational resources and time. Among classical decoders, the Union-Find (UF) decoder~\cite{delfosse2021almost, heer2023novel, liyanage2024fpga} is currently the fastest, leveraging the union-find data structure to achieve a worst-case runtime of $O(n \log n)$. However, its practical scalability is constrained by hardware limitations. On systolic arrays, it is restricted to code distances up to $d=15$~\cite{10196619}, while an FPGA implementation extends this to $d=51$ under 0.1\% phenomenological noise~\cite{liyanage2024fpga}. Despite these optimizations, existing decoders generally suffer from high hardware topological complexity, limiting their feasibility for scalability in large-scale FTQC\@.

Recently, an Ising model-based decoder with low topological complexity was proposed~\cite{fujisaki2022practical, fujisaki2023quantum}. This method treats data qubits as spins and syndrome values as coefficients, framing the QEC decoding task as an Ising model problem that can be solved using annealing machines~\cite{Fujitsu}, with scalability up to a code distance of 46. Although this method exhibits polynomial computational scaling $O(n^{1.01\sim1.84})$, which is slightly more efficient than original MWPM decoders, its decoding time remains suboptimal, and the logical error rate threshold is limited to a maximum of 9.8\%.

To address these limitations, we propose the Symmetric One-Hot Matching Elector (SOME), a novel decoder that first reformulates the QEC decoding problem as a Quadratic Unconstrained Binary Optimization problem---which we call the One-Hot QUBO (OHQ)---mathematically equivalent to an Ising model. In OHQ, each binary variable indicates whether a given pair of flipped syndromes is matched, and interaction coefficients encode the corresponding error probabilities. A one-hot constraint enforces that each flipped syndrome is matched exactly once, so valid solutions correspond to symmetric, self-inverse permutation matrices. Because OHQ turns the core matching problem into a standard QUBO format, it is solver-agnostic and can be addressed by any QUBO-capable method, including simulated annealing (SA), quantum annealing (QA), or other heuristics.

To fully leverage the structure of the solution space, which consists of self-inverse permutation matrices defined by one-hot encoding and symmetry, SOME reformulates decoding as the problem of finding the permutation matrix with the minimal total weight (interaction coefficients). It represents each flipped‑syndrome pair as a ``1” entry in a candidate matrix, sorts all pairs by weight, and uses each minimum‑weight pair to place the initial ``1”. Additional pairs are then appended in ascending order of weight to complete each candidate. By generating a diverse ensemble of candidates, SOME probes multiple low-energy regions of the solution space and greatly increases its chances of identifying the true global minimum. Finally, after computing the total energy of every candidate, it selects the permutation matrix with the lowest energy as the decoded solution.

To the best of our knowledge, SOME is the first decoder explicitly built for QEC decoding using binary self-inverse permutation matrices. This paper presents the following novel contributions.
\begin{itemize}
    \item Drastic variable count reduction: SOME reduces the number of variables by up to 99.9x compared to the state-of-the-art (SOTA) Ising model-based decoder~\cite{fujisaki2022practical}. For example, at a code distance of 100 and a 0.1\% physical error rate, the SOTA method requires an average of 58,807 variables, whereas SOME uses only 37.77.
    
    \item 
    Fast decoding on classical hardware: 
    Even on a classical CPU architecture, SOME completes decoding in microseconds, markedly faster than traditional decoders, which often require milliseconds to seconds for similar problem sizes.
    
    \item Improved logical error threshold: OHQ formulation attains a logical error threshold of 10.5\%, exceeding the best-known threshold of 10.3\% achieved by the MWPM decoder, confirming the robustness of the OHQ construction.
\end{itemize}

The rest of the paper is organized as follows: Sec. II briefly introduces  surface codes, MWPM and SOTA Ising models. Sec. III details our One-Hot QUBO (OHQ) construction. Sec. IV introduces the Symmetric One-Hot Matching Elector (SOME). Sec. V presents the experimental results, including the number of variables, decoding time, and the logical error rate threshold. Finally, Sec. VI concludes the paper.

\begin{figure}
 \vspace{-1.0em}
	\centering
	\includegraphics[width=1.\linewidth]{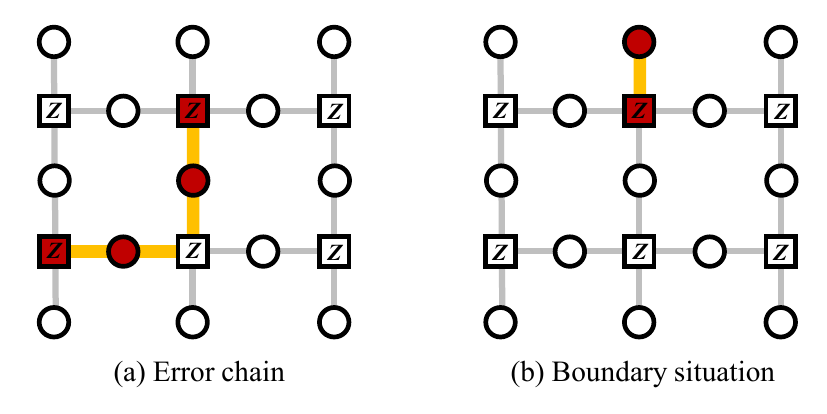}
	\caption{Error patterns. (a) Error chain: neighboring errors generate an error chain, and the syndrome flips at both ends of the error chain. (b) Boundary situation: The error in the first and last rows of the surface code will only flip one syndrome.}
	\label{fig:errorpattern}
\vspace{-1.0em}
\end{figure}

\section{Preliminaries}

\subsection{Surface Code and Quantum Error Correction}

This paper focuses on the planar surface code~\cite{fowler2012surface, kitaev2003fault}, which demonstrates high experimental feasibility. However, the proposed decoder is also applicable to other surface codes, such as the rotated surface code~\cite{horsman2012surface, bonilla2021xzzx} and the color code~\cite{bombin2006topological}. 

In QEC, as depicted in Fig.~\ref{fig:qec}~(a), the surface code consists of two types of qubits: data qubits (circles), and ancilla qubits (squares). These qubits are prone to errors induced by various noise sources, leading to bit-flip errors (X errors) or phase-flip errors (Z errors). Ancilla qubits detect errors in data qubits without direct measurement, with X-type ancilla qubits identifying Z errors and Z-type ancilla qubits detecting X errors. For code distance $d$, the number of data qubits is given by $d^2+(d-1)^2$, while both types of ancilla qubits number is $d \times (d+1)$. This arrangement ensures a Hilbert space dimension of 2, representing a logical qubit as a whole~\cite{devitt2013quantum}. Therefore, the Z ancilla qubits in the top and bottom rows, as well as X ancilla qubits in the left and right columns, are each connected to three data qubits.

Syndromes represent the collective logical state of data qubits as measured by ancilla qubits. Syndrome $+1$ indicates there is no error or an even number of errors in the surrounding data qubits, while $-1$ reflects an odd number of errors in the surrounding data qubits. The principle arises from the anti-commutation of Pauli operators X and Z, where $XZ+ZX=0$~\cite{nielsen2010quantum}. 

As shown in Fig.~\ref{fig:qec}~(b)--(d), an X error on a data qubit changes its representation from an open circle to a filled red circle. This flips the syndrome from $+1$ to $-1$, altering the corresponding ancilla qubit's symbol changes from an open square to a filled red square. The QEC decoder deduces error locations from these syndrome observations.

For simplicity, X and Z errors are assumed to occur independently with probability $p$, allowing them to be corrected separately. In this paper, we focus solely on X errors detected by Z ancilla qubits. The treatment of Z errors is entirely analogous, as rotating the surface code by 90 degrees swaps the positions of X ancilla qubits with those of Z ancilla qubits.

\subsection{Error Pattern}
An error in a data qubit causes the syndromes of two adjacent qubits to flip. When errors occur consecutively (i.e., in a continuous sequence in space) along data qubits, the syndrome flip is confined to the two endpoints of the error chain, forming a chain of errors (Fig.~\ref{fig:errorpattern}(a)). Any valid error chain connects flipped syndromes, as long as the chain does not cross the boundaries of the lattice. 

A special case arises in situations involving boundary conditions, where errors in data qubits located on the first or the last row flip only a single syndrome (Fig.~\ref{fig:errorpattern}(b)). 

In these cases, the flipped syndromes may form chains either by connecting to other syndromes or extending to the boundary. Thus, the QEC problem can be reformulated as a graph-theoretic matching problem, where syndromes must be matched to form valid error chains or, in special cases, remain unmatched due to boundary constraints.

\subsection{Minimum-Weight Perfect Matching for QEC}
MWPM constructs a graph where each vertex corresponds to a flipped syndrome. It connects every vertex to all others, assigning edge weights based on the minimum distance between the corresponding syndrome positions on the lattice. However, boundary conditions introduce additional complexity. To handle such situations, MWPM employs the Blossom algorithm~\cite{edmonds1973matching}, which handles complex matching scenarios by grouping odd-length cycles of tight edges---known as blossoms---into single nodes. When an odd number of vertices are connected in such a cycle, or when existing blossoms form new cycles, they are recursively collapsed into new blossoms, enabling the algorithm to proceed efficiently with alternating trees and matchings.

\subsection{Ising Models for QEC}

The Ising model-based decoder~\cite{fujisaki2022practical} 
transforms the QEC into the following Ising model: 
\begin{equation}
    H = -J\sum_v^{N_v}b_v\prod_{i\in\delta v}^{4\,\mathrm{or}\,3}\sigma_i-h\sum_i^{N_d}\sigma_i. \notag
\end{equation}
Data qubits are represented by variables $\sigma_i$, which flip from $+1$ to $-1$ during the annealing process if an error occurs. An ancilla qubit is denoted by $v$, with $\delta v$ representing the list of connected qubits.
The syndrome $b_v$ is $+1$ for even or no errors and $-1$ for odd errors. $N_v$ and $N_d$ represent the number of syndromes and data qubits, respectively. The first term, with coupling coefficient $J$, enforces consistency between the decoded error pattern and the observed syndromes. The second term, with external field coefficient $h$, minimizes the total errors. 
Since each data qubit can independently take one of two spin states ($+1$ or $-1$), the total solution space has a size of $2^{N_d}$. This exponential growth makes finding the solution of the system computationally challenging.

\begin{figure}
	\centering
	\includegraphics[width=1.\linewidth]{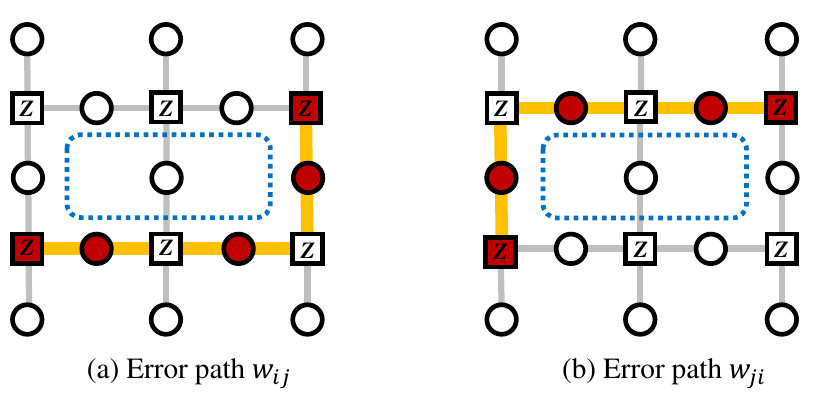}
	\caption{Trivial loop (blue circle): a trivial loop corresponds to a closed error path that does not cross any logical operator. Different correction paths within the loop are equivalent, as they yield the same syndrome and have no logical effect.}
	\label{fig:trivial_loop}
\vspace{-1.0em}
\end{figure}
\section{One-Hot QUBO Formulation}
\label{ISM}
In this section, we propose a novel QUBO formulation referred to as One-Hot QUBO (OHQ) for QEC problems. The matching between pairs of flipped syndromes is represented using binary variables, forming a syndrome matching matrix. Boundary situations are addressed in a particularly efficient manner, as self-matching, which is embedded into the diagonal entries of the matrix, allowing them to be handled without introducing additional graph structures.

Each pair of flipped syndromes may be connected through an error chain, and the number of physical errors, together with the associated error probabilities, is used to construct a weight matrix. To ensure that each flipped syndrome is matched exactly once, either to another syndrome or to the boundary, a set of one-hot constraints is introduced. The QUBO is obtained by performing an element-wise multiplication between the syndrome matching matrix and the weight matrix, with an additional penalty term for constraint violation. By minimizing this QUBO energy, an error configuration with the fewest corrections is identified.

\subsection{Syndrome Matching Matrix}
When there are $n$ flipped syndromes, they form a $n\times n$ matrix $X = [x_{ij}]$, 
where each element $x_{ij}$ is binary variable. When $i \ne j$, the element represents a matching between two distinct syndromes, while $i=j$, corresponds to a self-matching case, which models the syndrome being matched to the boundary.

Traditionally, the syndrome matching matrix is assumed to be symmetric, i.e., $x_{ij} = x_{ji}$, under the assumption that error chains forming a trivial loop (see Fig.~\ref{fig:trivial_loop}) are physically indistinguishable. Consequently, the syndrome matching matrix in OHQ is constructed with symmetry across the main diagonal, as illustrated:
\[
X = 
\begin{bmatrix}
x_{00} & x_{01} & x_{02} & \cdots & x_{0(n{-}1)} \\
x_{01} & x_{11} & x_{12} & \cdots & x_{1(n{-}1)} \\
x_{02} & x_{12} & x_{22} & \cdots & x_{2(n{-}1)} \\
\vdots & \vdots & \vdots & \ddots & \vdots       \\
x_{0(n{-}1)} & x_{1(n{-}1)} & x_{2(n{-}1)} & \cdots & x_{(n{-}1)(n{-}1)} \\
\end{bmatrix}
\]
\subsection{Weight Matrix}

The weight matrix is denoted as $W = [w_{ij}]$, where each $w_{ij}$ represents the cost associated with matching flipped syndromes $i$ and $j$. If the error chain connecting syndromes $i$ and $j$ involves $k$ physical errors, the associated probability is given by $P = \prod_i^k p_i$. As each error is assumed to occur with identical probability $p$, the expression simplifies to $P = p^k$. By taking the logarithm, the value becomes $P = k\log p$. Since $\log p$ is a constant across all terms, it can be eliminated without affecting the optimization, and the weights $w_{ij}$ can therefore be defined directly in terms of the number of errors $k$. For example, the error chains shown in Fig.~\ref{fig:errorpattern}~(b), Fig.~\ref{fig:errorpattern}~(a), and Fig.~\ref{fig:trivial_loop}(a) correspond to $k=1, 2$ and $3$, respectively.

To account for symmetry in the matrix, the weights are adjusted as follows: if $i=j$, corresponding to a self-match (i.e., a match to the boundary), the full error count is assigned, and $w_{ii} = k$. If $i\ne j$, the weight is shared between symmetric entries, and each off-diagonal element is assigned $w_{ij} = \frac{k}{2}$. This ensures that each error chain contributes equally to the total energy.

Notably, if the number of errors between two flipped syndromes exceeds the sum of their respective distances to the boundary, this indicates the formation of a nontrivial error loop, and the decoded result may alter the logical qubit state. As such cases fall outside the correctable range, they cannot be reliably resolved through annealing. Specifically, when the error count $k$ between two syndromes exceeds $\frac{d-1}{2}$, the corresponding elements in both the syndrome matching matrix and the weight matrix are set to zero, effectively excluding these matches from the optimization process.

However, it is noted that in OHQ method, the symmetry constraint is relaxed. The matrix $W$ is allowed to be asymmetric, enabling the model to capture directional information in the error paths. That is, $w_{ij}$ and $w_{ji}$ represent two distinct decoding paths. This distinction is especially useful when the error path includes physical directionality~\cite{bohm1999time} (e.g., different lengths, timing, or asymmetric noise), which cannot be captured by symmetric formulations.

\subsection{Constraint}
Since each flipped syndrome is required to be uniquely matched exactly once, either with itself or with another flipped syndrome, a constraint is imposed for each row of the matching matrix. Specifically, the sum of matchings associated with each flipped syndrome must equal 1, that is, {$\sum_{j=1}^n x_{ij}$} = 1. 
In the QUBO model, this constraint is reformulated in a squared form to penalize violations. Accordingly, the total constraint term for all flipped syndromes is expressed as:
\begin{equation}
    \text{constraint} = \sum_{i=1}^n(\sum_{j=1}^nx_{ij} - 1)^2. \notag
\end{equation}
\subsection{Problem Formulation}
When constructing the QUBO model, element-wise Hadamard multiplication is performed between the syndrome matching matrix $X$ and the weight matrix $W$ to compute the objective energy. In addition, a constraint term is incorporated for each flipped syndrome to ensure one-hot matching. The resulting QUBO formulation is given by:
\begin{equation}
    H = X \odot W + \text{penalty}\times\sum_{i=1}^n(\sum_{j=1}^nx_{ij} - 1)^2. \notag
\end{equation}
Here, penalty denotes the penalty weight assigned to the constraint term, ensuring that constraint violations are discouraged during the optimization process. Since any correctable error configuration must involve at most $\frac{d-1}{2}$ errors, the penalty is chosen to be sufficiently large, and is specifically set to $\text{penalty}=d^2$ in this work. As each weight $w_{ij}$ represents the number of physical errors in the corresponding error chain, minimizing the QUBO energy directly corresponds to minimizing the total number of errors---thereby yielding the most likely error configuration under the given syndrome.

\subsection{Error Correction}
Once the QUBO model is processed by the annealer, a result of $x_{ij} =1$ indicates a match between the $i$-th and $j$-th flipped syndromes. The data qubits along the shortest path between these two syndromes are then identified for correction. As noted in Sec.~3.1, error paths forming a trivial loop are regarded as equivalent, and thus the correction path is selected at random---either proceeding horizontally first and then vertically, or vice versa.

\begin{algorithm}[t]
\caption{Greedy Multi-Seed Permutation Vector Generation}\label{SOME}
\KwIn{Weight matrix $W\in\mathbb{R}^{n\times n}$}
\KwOut{Optimized permutation vector $V^*$}

Let $\mathit{P}=\{(i,j,W_{ij}) : 0\le i\le j<n\}$\;
Sort $\mathit{P}$ by weight ascending; off-diagonal before diagonal\;
Let $w_{\min}=\mathit{P}[0].w$, and let $k=\#\{p\in\mathit{P}\mid p.w=w_{\min}\}$\;
$\mathrm{BestEnergy}\gets+\infty$\;
\For{$s=0$ \KwTo $k-1$}{
  Initialize $V_{\rm tmp}$ empty, $\mathrm{Used}=\emptyset$\;
  Seed $(r,c)\gets\mathit{P}[s]$,\\
  set $V_{\rm tmp}[r]=c$, $V_{\rm tmp}[c]=r$, add $\{r,c\}$ to Used\;

  \ForEach{$(i,j)\in\mathit{C}$}{
    \If{$i,j\notin\mathrm{Used}$}{
      $V_{\rm tmp}[i]=j,\;V_{\rm tmp}[j]=i$, add $\{i,j\}$ to Used\;
    }
  }
Let Diags = $\{i:V_{\rm tmp}[i]=i\}$\;
  \If{$|\mathrm{Diags}|\ge2$}{
    Swap any pair in Diags if it lowers total energy\;
  }

  $E\gets\sum_i W_{i,V_{\rm tmp}[i]}$\;
  \If{$E<\mathrm{BestEnergy}$}{
    $\mathrm{BestEnergy}\gets E$, $V^*\gets V_{\rm tmp}$\;
  }
}

\Return $V^*$
\end{algorithm}

\section{Symmetric One-hot Matching Elector}
To further improve the efficiency of decoding, we propose a novel decoder named the Symmetric One-hot Matching Elector (SOME). This method is specifically designed to exploit the structural properties of the solution space in the QUBO formulation, where valid solutions are restricted to self-inverse permutation matrices---matrices that are both one-hot and symmetric with respect to the main diagonal.

SOME reformulates decoding as the task of identifying the permutation matrix with the smallest total weight. It begins by ranking every flipped-syndrome pair by its weight and using each of the smallest weight pairs to seed a candidate matrix with an initial ``1.” The algorithm then fills in the remaining ``1” entries by greedily adding pairs in order of increasing weight until each matrix is complete. By producing a varied set of candidates from different seeds, SOME samples multiple low-energy configurations and boosts its chance of finding the true global minimum. After computing the total energy for each candidate, the algorithm selects the permutation matrix with the lowest energy as its final decoded solution.

\subsection{Self-inverse Permutation Matrices}
A self-inverse permutation matrix\footnote{also known as Involutive Matrix} is a special class of permutation matrices that satisfies two structural properties: (i) it is symmetric with respect to the main diagonal, and (ii) each row and each column contains exactly one entry equal to 1. These properties imply that such a matrix is equal to its inverse, i.e., $P=P^{-1}$, and also symmetric, i.e., $P=P^{T}$.

Due to this structure, a self-inverse permutation matrix can be compactly represented by a one-dimensional vector $V$, which we refer to as the permutation vector. Each element of this vector encodes a matching pair in the matrix. Specifically, for each index $x$, the value $V[x] = y$ indicates that the matrix has 1 at both positions $(x,y)$ and $(y,x)$, implying symmetry and mutual assignment. This also means that $V[y] = x$, i.e., the relation is involutive. Diagonal elements (i.e., $V[x] =x$) represent self-matching positions, typically used to model boundary cases.
\subsection{Multi-Seed Permutation Vector Generation}

We now describe the SOME algorithm, the complete pseudocode is shown in Algorithm~\ref{SOME}. First, we enumerate every unordered syndrome pair $(i,j)$ along with its associated weight $W_{ij}$ into a flat pair set $P$. This set is then stably sorted in ascending order according to weight, with ties broken in favor of off-diagonal pairs to encourage nontrivial matchings early on. Let $w_{min}$ denote the smallest weight in $P$ and let $k$ be the count of entries achieving $w_{ij}$; these $k$ entries become the ``seeds.'' For each seed $(r, c)$, we initialize a temporary permutation vector $V_{tmp}$ by pairing $r$ with $c$ and vice versa, marking both indices as used. We then greedily append the next lowest-weight unused pair from $P$, iterating until all indices are assigned, guaranteeing a full one-hot permutation vector. 


Optionally, if two or more self-paired (diagonal) entries remain, we attempt pairwise swaps among them whenever such a swap reduces the total weight. Each completed vector thus corresponds to a heuristic matching built from a different low-weight starting point. Finally, we compute the total weight $\sum_iW_{i,V_{tmp}[i]}$ for each of the $k$ seeds and select the one with the smallest sum as the decoded solution. Combining a multi-seed greedy strategy with randomized swaps broadens the search across multiple low-energy regions of the solution space.

\begin{table*}[t]
\centering
\caption{Variable number comparison between SOTA, OHQ-AE and SOME.}
\label{table: Table 1}
\resizebox{2.\columnwidth}{!}{
\Huge
\renewcommand{\arraystretch}{1.5}
\begin{tabular}{@{}ccc|c|ccccccc|ccccccc@{}}
\toprule
\multirow{2}{*}{$d$} & \multirow{2}{*}{\#data qubits} & \multirow{2}{*}{\#syndromes} & \multirow{2}{*}{\#\cite{fujisaki2022practical}} & \multicolumn{7}{c|}{\#OHQ-AE with physical error rate (\%)}  & \multicolumn{7}{c}{\#SOME with physical error rate (\%)}  \\ \cmidrule(l){5-18} 
   &       &       &       & 0.1 & 1     & 2      & 5       & 10     & 15       & 20     & 0.1  & 1    & 2     & 5     & 10    & 15   & 20  \\ \midrule
5  & 41    & 30    & 97   & 0.08 & 1.21  & 1.72   & 5.02   & 12.15  & 18.25     & 22.12  & 0.07 & 0.75 & 1.13  & 2.83  & 5.52  & 7.18 & 8.3  \\
9  & 145   & 90    & 385  & 0.38 & 4.6   & 13.27  & 40.46  & 95.62 & 160.67   & 210.36   & 0.23 & 2.45 & 5.3   & 11.43 & 19.5  & 26.7 & 30.81 \\
13 & 193   & 156   & 865  & 0.61 & 16.48 & 39.16  & 157.26  & 427.06  & 675.54 &881.25   & 0.41 & 6.24 & 11.08 & 25.79 & 44.98 & 57.6 & 66.76 \\
17 & 545   & 272   & 1537 & 1.22 & 30.21 & 89.08  & 427.06  & 1201.10  &1870.46  &2592.97 & 0.80  & 9.62 & 18.73 & 45.07 & 77.75 & 98.67& 117.18 \\
21 & 841   & 420   & 2401 & 2.67     & 64.51 & 210.78 & 985.97 & 2817.38   &4567.8    &5972.89   & 1.58 & 15.5 & 30.52 & 70.18 & 121.57& 156.45 & 179.45\\
25 & 1201  & 600   & 3457 & 4.85    & 123.69      & 416.84      & 1934.37       & 5740.07      &9211.23        & 12158.07      & 2.57 & 22.35 & 44.68 & 100.48& 175.37& 224.33 & 257.99\\  \midrule
50 & 4901  & 2450  & 14407  & 24.92   & 1714.90      & 5601.54      & 30893.01       & 88411.98      & 147005.81   & 192684.67   & 8.70  & 96.54 & 177.95& 422.41& 718.68& 926.96 & 1062.22\\
100 & 19801  & 9900&58807      & 298.47   & 25670.44      & 95793.75      & 495319.77       & 1455120.20      & 2402406.36    & 3173007.80      & 37.77& 382.36& 744.29& 1697.31& 2914.77 & 3746.57& 4309.70\\
\bottomrule
\end{tabular}
}

\end{table*}

\begin{table}[t]
\centering
\caption{The decoding time for 100 error patterns using AE for SOTA and the proposed OHQ}
\label{table: Table 7}
\resizebox{1.\columnwidth}{!}{
\begin{tabular}{@{}c|ccccc|ccccc@{}}
\toprule
\multirow{2}{*}{d} & \multicolumn{5}{c|}{SOTA (ms)}   & \multicolumn{5}{c}{OHQ (ms)}    \\ \cmidrule(l){2-11} 
                   & 1\% & 5\% & 10\%  & 15\% & 20\%  & 1\% & 5\% & 10\%  & 15\% & 20\% \\ \midrule
5                  & 84   & 80   & 83   & 90   & 135  & 27   & 29   & 34  & 39   & 42  \\
9                  & 91   & 1187 & 1906 & 1676 & 1947 & 30   & 51   & 60  & 58   & 57  \\
13                 & 411  & 3010 & 2919 & 2608 & 2794 & 39   & 58   & 62  & 96   & 153 \\
17                 & 1594 & 3405 & 3514 & 3395 & 3576 & 47   & 60   & 147 & 196  & 277 \\
21                 & 2836 & 4091 & 4060 & 3882 & 4386 & 59   & 110  & 264 & 599  & 812 \\ \bottomrule
\end{tabular}
}
\end{table}

\begin{table}[t]
\centering
\caption{The average decoding time for 100 error patterns (without error-free pattern) using quantum annealer D-Wave to solve OHQ}
\label{table: Table 4}
\resizebox{1.\columnwidth}{!}{
\setlength{\tabcolsep}{15pt} 
\renewcommand{\arraystretch}{1} %
\begin{tabular}{@{}c|cc@{}}
\toprule
\multirow{2}{*}{\quad$d$} & \multicolumn{2}{c}{Decoding time under the practical physical error rate  ($\mu$s)} \\ \cmidrule(l){2-3} 
                     & \multicolumn{1}{c|}{\quad\quad\quad\quad\quad\quad0.1\%\quad\quad\quad\quad\quad\quad}    & 1\%     \\ \midrule
\quad5                    & \multicolumn{1}{c|}{\quad\quad\quad20\quad\quad\quad\quad}     & 36    \\
\quad7                    & \multicolumn{1}{c|}{\quad\quad\quad30\quad\quad\quad\quad}     & 50    \\
\quad9                    & \multicolumn{1}{c|}{\quad\quad\quad70\quad\quad\quad\quad}     & 101    \\ \bottomrule
\end{tabular}
}
\end{table}
\section{Numerical Experiments}

The experiments were conducted using Python~3.11.9 on a classical computer with a Core i5-13400 CPU and 16\,GB of memory for OHQ construction. The model was solved using the following solvers: simulated quantum annealing (SQA) via a GPU-accelerated cloud server---Amplify Annealing Engine (AE)\cite{Fixstars}, real quantum annealing (QA) through the D-Wave cloud server~\cite{Dwave}, locally deployed SOME. The SOME was implemented in C++ for performance optimization and called as a library from python scripts.

\subsection{Number of Variables}
\subsubsection{Data Setup}
\indent
We compared the number of variables in OHQ solved by AE and SOME with the SOTA Ising model-based decoder solved by AE~\cite{fujisaki2022practical} across code distances ranging from 5 to 25, as well as 50 and 100. For each distance, 100 random error patterns were tested under physical error rates of 0.1\%, 1\%, 2\%, 5\%, 10\%, 15\% and 20\%. We convert the cubic term to a quadratic form by introducing an auxiliary variable $y$ and a penalty term, using the identity $x_1x_2x_3 = yx_3 + \text{Penalty} \cdot (x_1x_2-2x_1y-2x_2y+3y)$ as proposed in~\cite{glover2022quantum}. In SOTA, the variable count remains constant for a given code distance, as all syndromes and data qubits are considered, with each represented as an Ising model variable. In contrast, OHQ defines variables only for flipped syndromes, where each variable indicates whether a pair of flipped syndromes is matched. Consequently, the total number of variables in OHQ and SOME varies depending on the specific error pattern and physical error rate.

Table~\ref{table: Table 1} presents the code distance ($d$), the number of data qubits, the number of syndromes, the constant variable count for the Ising model-based decoder~\cite{fujisaki2022practical}, and the average variable count for OHQ solved by AE and SOME under different error rates.

\subsubsection{Performance Analysis}
\indent
The OHQ solved by AE, which is called OHQ-AE, significantly reduces the required variables compared to SOTA~\cite{fujisaki2022practical} at low physical error rates. As the code distance increases and physical error rates become higher, OHQ-AE's performance lags behind that of SOTA. However, within practical error ranges ($<1$\%), OHQ-QE is much faster than the SOTA. In contrast, the SOME further consistently reduces the required variable count. 

OHQ-AE outperforms SOTA at low and practical physical error rates, but the variable count increases rapidly. The SOTA constructs the Ising model by including all data qubits and introduces third- and fourth-order terms to represent interactions around each syndrome. These higher-order terms introduce additional variables, and can result in the total number of variables exceeding the number of data qubits, thereby significantly increasing computational complexity. The overall growth follows an $O(n)$ scaling with respect to the number of qubits. In contrast, OHQ-AE only considers flipped syndromes when constructing the QUBO model, and at low physical error rates, the number of flipped syndromes is relatively small. Furthermore, the terms in the QUBO model are no higher than quadratic, avoiding the introduction of unnecessary variables. However, at high physical error rates, as the number of flipped syndromes increases substantially, the matrix structure becomes $O(n^2)$, leading to a quadratic growth in the number of variables. However, it is important to note that QUBO solving time does not scale purely with variable count but also with problem complexity. In SOTA, variable interactions are represented by higher-order terms, which will create a rugged energy landscape in which global minima are hard to locate. In addition, the SOTA can produce invalid solutions, which further enlarges the solution space to be explored and makes it difficult to find the best solution. In contrast, OHQ restricts all interaction terms to at most quadratic and encodes only binary “matched or not” decisions for each syndrome pair, producing a much simpler QUBO structure. As a result, even when a SOTA formulation uses fewer variables than OHQ, it may still require longer solve times and face greater difficulty in reaching the true global optimum.

SOME consistently outperforms the SOTA by reducing the problem dimensionality from a matrix to a vector, resulting in a linear variable growth of $O(n)$. The ``$n$" in SOME scales with the number of flipped syndromes, which enables SOME to achieve a significant reduction in variable count across a wide range of scenarios.

To further validate its scalability, the number of variables for code distances $d = 50$ and $d = 100$ are evaluated. At a 0.1\% physical error rate, SOME required only an average of 37.77 variables for $d = 100$, demonstrating its efficiency even at large code distances. Given that current advanced quantum hardware achieves physical error rates in the range of 0.1\%~\cite{google2023suppressing} to 1\%~\cite{fowler2009high}, and error rates are expected to continue to decrease, SOME is well suited for practical deployment in large-distance surface codes under realistic noise levels.

\subsection{Decoding Time}
Decoding time is critical, since qubit states may collapse before the decoding process completes if it takes too long. Using the AE annealer, we measured the time required by both the SOTA and OHQ to reach the global optimum---or, for the most challenging instances, a local optimum. Experiments were performed for code distances $d$ from 5 to 21 under physical error rates of 1\%, 5\%, 10\%, 15\%, and 20\%. For each ($d$, physical error rate) pair, we generated 100 random error patterns (excluding those with no errors) and report the average decoding time in Table~\ref{table: Table 7}. Up to 30x faster decoding speed is observed for OHQ compared to the SOTA. However, despite delivering a multi-fold acceleration over the SOTA decoder, the decoding time remains on the order of milliseconds, which is still suboptimal.

Due to current constraints of the D-Wave quantum annealer on the maximum QUBO size, only the average annealing time for 100 error patterns at physical error rates of 0.1\% and 1\% was measured, as summarized in Table~\ref{table: Table 4}. The results reveal that, when considering decoding time alone, OHQ attains microsecond-level performance. While this evaluation underscores the method’s potential, it must be recognized that the communication latency between classical host systems and D-Wave’s cloud servers incurs a substantial overhead. Consequently, a rapidly executable, locally operable QEC solution such as SOME is imperative.

\begin{figure}
	\centering
	\includegraphics[width=.95\linewidth]{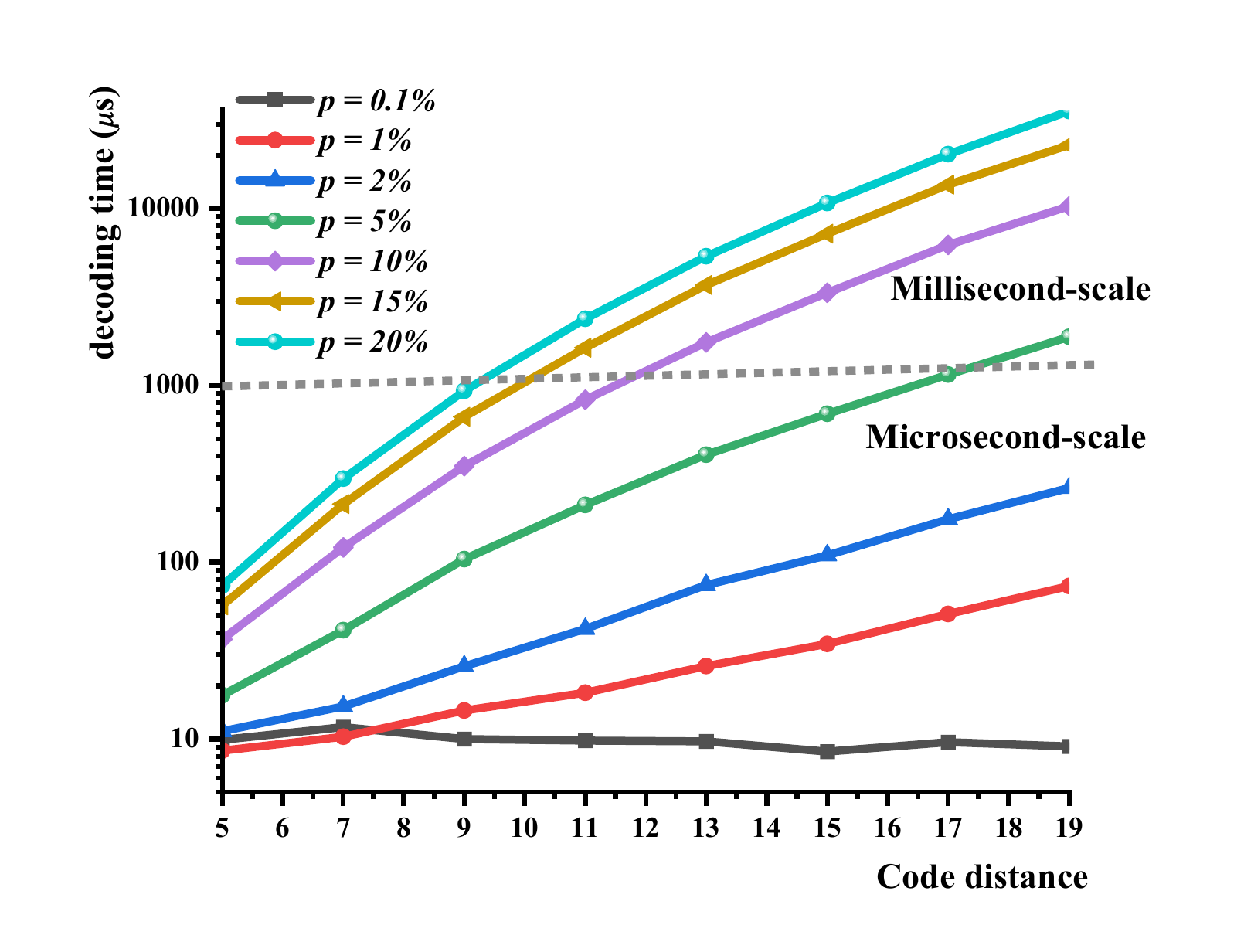}
	\caption{SOME’s average decoding times for code distances from $d$ = 5 to 19 at physical error rates of 0.1\%, 1\%, 2\%, 5\%, 10\%, 15\% and 20\%.}
	\label{fig:decoding_time}
\end{figure}

We evaluated 10000 error patterns (none error‑free) for code distances $d=5$ to 19 at physical error rates of 0.1\%, 1\%, 2\%, 5\%, 10\%, 15\% and 20\%, and measured the average decoding time (see Fig.~\ref{fig:decoding_time}). At 0.1\% and 1\% error rates, decoding completes in just a few microseconds; as both the error rate and the code distance increase, the decoding time rises sharply. Nevertheless, even at a 20\% physical error rate with $d=19$, decoding remains in the millisecond regime. Although all measurements reported here were performed on a single commodity CPU thread, SOME can be easily parallelized. Spreading different seeds over multiple CPU cores or offloading the task to a GPU/FPGA is expected to deliver significant additional performance improvements in terms of speed and solution quality.
\subsection{Logical Error Rate}

Logical error rate is another key metric for evaluating QEC decoders. We assessed the threshold behavior of the proposed OHQ method on an AE annealer, as well as that of the SOME decoder, across code distances from 5 to 25. The threshold is defined as the physical error rate below which the logical error rate decreases as the code distance increases. For each physical error rate, we tested 10,000 random error patterns to estimate the logical error rate.

\begin{figure}
	\centering
	\includegraphics[width=.95\linewidth]{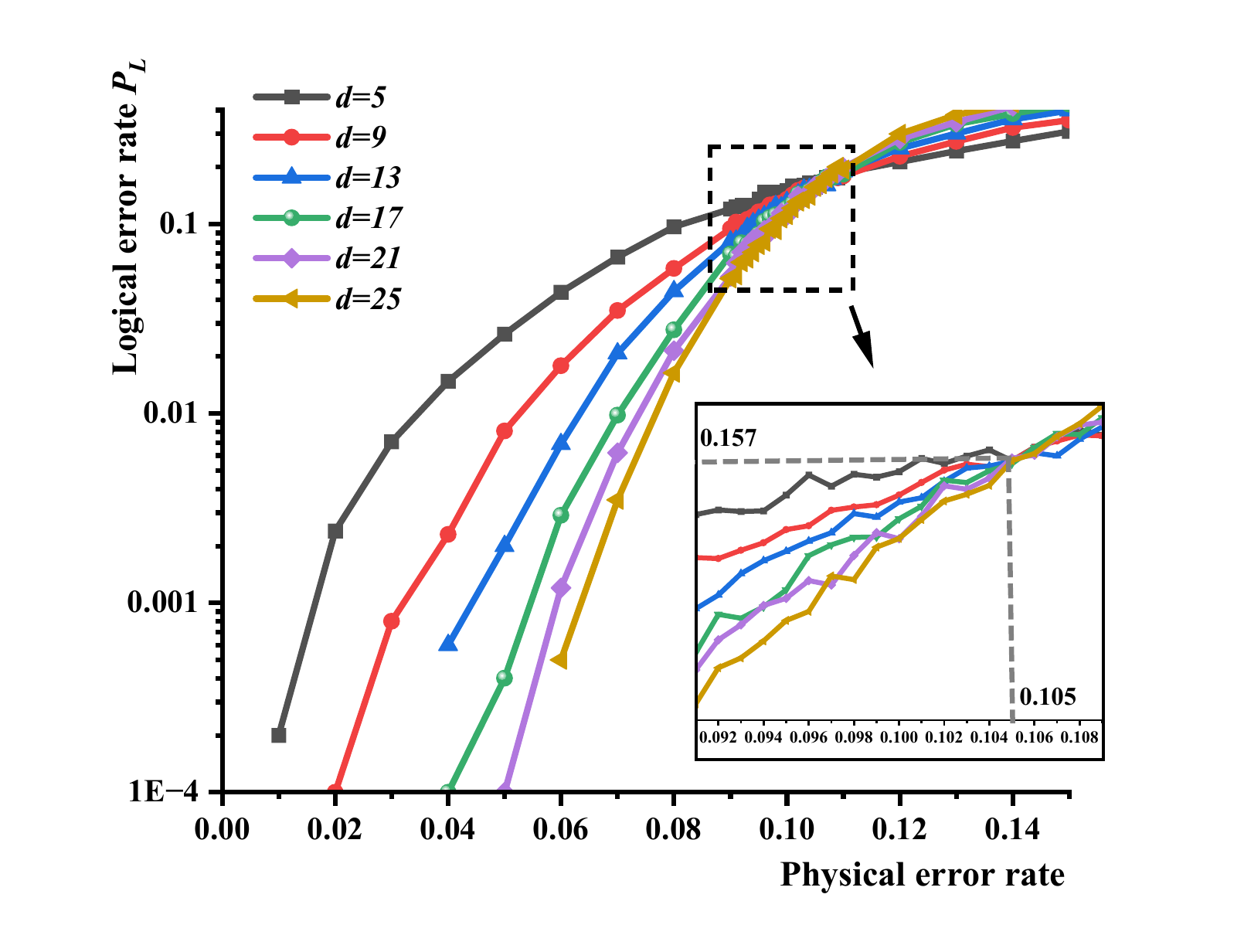}
	\caption{Correlation between the calculated logical error rate $P_L$ and the physical error rate $p$ for $d$ from 5 to 25 of OHQ-AE.}
	\label{fig:logical_error_rate}
\end{figure}

As shown in Fig.~\ref{fig:logical_error_rate}, because the decoding process is inherently stochastic, the logical error rate curves exhibit fluctuations rather than smooth trends. Nevertheless, all curves intersect at approximately a 10.5\% physical error rate, indicating an OHQ threshold of 10.5\%.

This value is slightly higher than the best-known threshold achieved by the MWPM decoder. The improved threshold of OHQ can be attributed to its ability to explore a broader solution space through annealing-based optimization, allowing it to correct certain error configurations that MWPM may misidentify, especially near the threshold region. Moreover, OHQ is not limited to strict pairwise matching and can flexibly model complex error correlations, which may contribute to its higher threshold. This confirms the strong error-suppression capability of OHQ formulation.

\begin{figure}
	\centering
	\includegraphics[width=.95\linewidth]{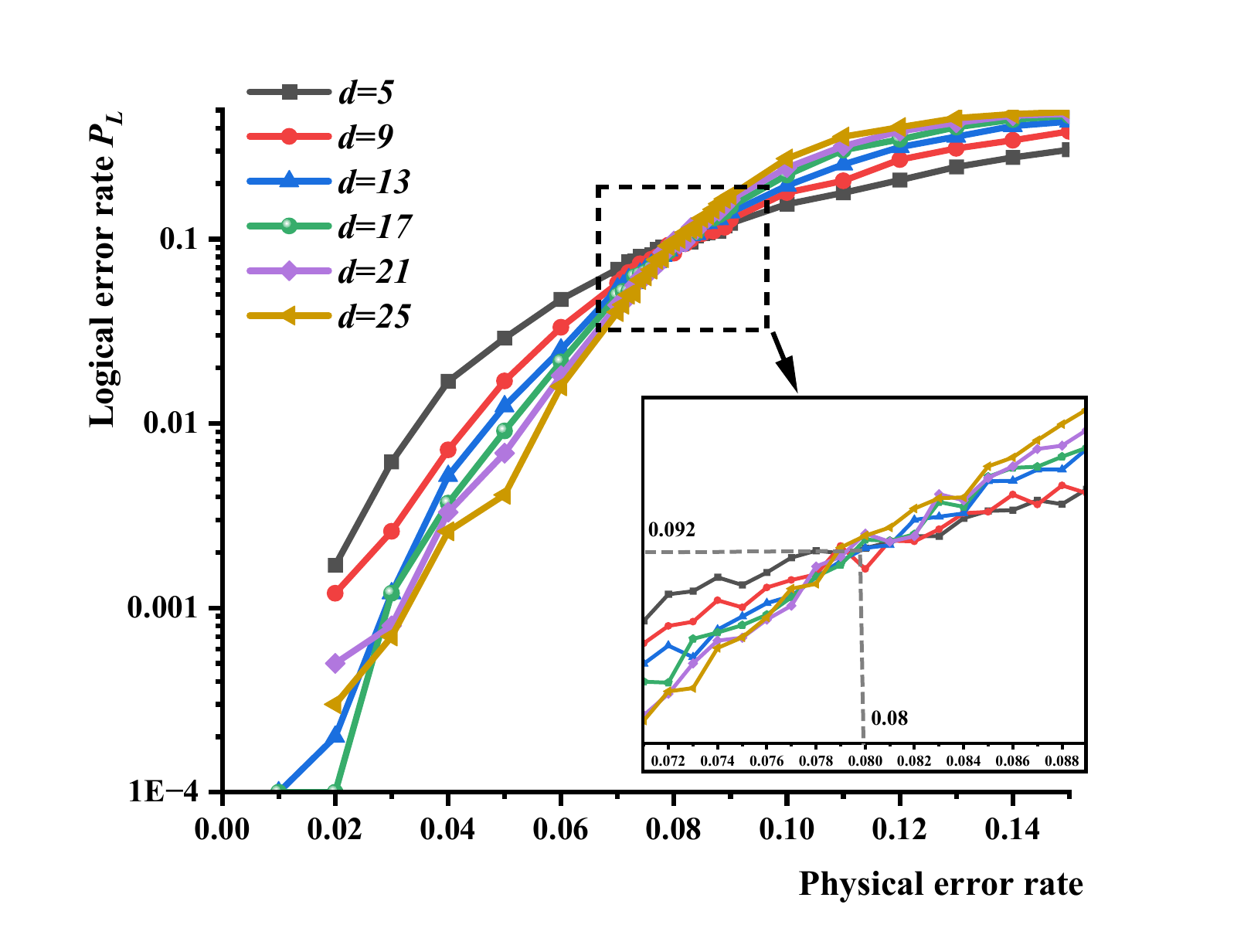}
	\caption{Correlation between the calculated logical error rate $P_L$ and the physical error rate $p$ for $d$ from 5 to 25 of SOME.}
	\label{fig:logical_SOME}
\end{figure}

Fig.~\ref{fig:logical_SOME} presents the threshold behavior of the SOME decoder. Because SOME restricts its matching to a greedy subset of low-weight syndrome pairs rather than exhaustively exploring all pairings, its logical error curves intersect at a slightly lower physical error rate of 8\%, which is sufficiently high for the current qubit error rates. However, it is worth noting that according to the similarities of the problem setting, by leveraging Traveling Salesman Problem (TSP)-inspired optimization techniques---such as applying genetic algorithms~\cite{fu2018solving} or greedy heuristics~\cite{gutin2002traveling}---one can more effectively approach the global optimum. In this work, we limit ourselves to proposing a new formulation for QEC decoding problems that avoids the need for extreme hardware resources while still providing microsecond decoding. Consequently, our future work includes the development of a fast and accurate OHQ solver to increase the logical error threshold even further.

\section{Conclusion}
In this paper, we propose the Symmetric One-Hot Matching Elector (SOME), a novel decoder that reformulates the QEC decoding problem as a simple Quadratic Unconstrained Binary Optimization (QUBO) problem---termed the One-hot QUBO (OHQ)---equivalent to an Ising model. In OHQ, each binary variable denotes whether a specific pair of flipped syndromes is matched, and interaction coefficients encode the corresponding error probabilities. A one-hot constraint ensures each syndrome is paired exactly once, so valid solutions are symmetric, self-inverse permutation matrices.

SOME exploits one‑hot encoding and the symmetry of self‑inverse permutation matrices to recast decoding as the search for the permutation matrix with the smallest total interaction weight. It begins by seeding each candidate matrix with a lowest‑weight flipped‑syndrome pair (encoded as a ``1'' entry) and then greedily adds remaining pairs in order of increasing weight until the matrix is complete. This process generates a diverse ensemble of candidates that probe multiple low‑energy regions of the solution space. By evaluating the total energy of each candidate and selecting the matrix with the lowest energy, SOME achieves fast decoding with minimal computational overhead.

On classical CPU hardware, SOME consistently completes decoding in  microseconds. Moreover, OHQ exhibits a threshold of 10.5\%, comparable to the MWPM decoder. This work establishes a new computational paradigm for QEC decoding that eschews the need for specialized, high-performance hardware while still delivering microsecond-level decoding performance.

\section*{Acknowledgment}

This work was partially supported by Japan Science and Technology agency (JST) Moonshot R\&D Grant Number JPMJMS226A.
It was also supported by JSPS KAKENHI grant Nos.~23K28052 and 23K18462. Additional support was provided by JST SPRING, Grant Number JPMJSP2110.

\bibliographystyle{unsrt}
\bibliography{main}

\end{document}